\documentclass[a4paper,10pt,twoside]{cpc-hepnp}

\usepackage{multicol}
\usepackage{graphicx}
\usepackage{booktabs}
\usepackage{amssymb,bm,mathrsfs,bbm,amscd}
\usepackage[tbtags]{amsmath}
\usepackage{lastpage}
\usepackage{CJK}
\usepackage{color}

\begin{document}
\begin{CJK*}{GBK}{song}

\fancyhead[c]{\small } \fancyfoot[C]{\small -\thepage}

\footnotetext[0]{}

\title{Two-Neutron Halo State of ${}^{15}$B Around $3.48$ MeV By A Three-Body Model}

\author{%
      Dong Bai$^{1;1)}$\email{dbai@itp.ac.cn}%
\quad Zhongzhou Ren$^{2;2)}$\email{Corresponding Author: zren@nju.edu.cn}%
\quad Tiekuang Dong$^{3;3)}$\email{tkdong@pmo.ac.cn}
%\quad F. Jone$^{2}$
}
\maketitle

\address{%
$^1$ School of Physics, Nanjing University, Nanjing, 210093, China\\
$^2$ School of Physics Science and Engineering, Tongji University, Shanghai 200092, China\\
$^3$ Purple Mountain Observatory, Chinese Academy of Sciences, Nanjing 210008, China
}

\begin{abstract}
We investigate low-lying bound states of the neutron-rich nucleus ${}^{15}$B by assuming it is a three-body system made of an inert core ${}^{13}$B and two valence neutrons. The three-body wave functions are obtained using the Faddeev formalism. Special attention is paid to the excited state at $3.48(6)$ MeV observed in the ${}^{13}\text{C}({}^{14}\text{C},{}^{12}\text{N}){}^{15}\text{B}$ reaction, whose properties are less clear theoretically. In our three-body model, besides the ground state $3/2_1^-$, a second $3/2_2^-$ state is discovered at around $3.61$ MeV, which might be identified with the excited state observed at $3.48(6)$ MeV. We study this $3/2_2^-$ state in detail, which turns out to be a two-neutron halo state with a large matter radius $r_\text{m}\approx 4.770$ fm.
\end{abstract}



%\footnotetext[0]{\hspace*{-3mm}\raisebox{0.3ex}{$\scriptstyle\copyright$}2013
%Chinese Physical Society and the Institute of High Energy Physics
%of the Chinese Academy of Sciences and the Institute
%of Modern Physics of the Chinese Academy of Sciences and IOP Publishing Ltd}%

\begin{multicols}{2}

\section{Introduction}

 Neutron-rich nuclei have attracted much attention in the last decades, and many novel structures are expected in these nuclei, such as { neutron halos \cite{Jensen:2004zz,Meng:1996tt,Terasaki:2006tw,Grasso:2006es}}, Borromean structures, Efimov-like systems \cite{Jensen:2004zz}, two-neutron emissions \cite{Pfutzner:2011ju}, etc. In this letter, we would like to study theoretically the neutron-rich nucleus ${}^{15}$B. Although discovered more than fifty years ago \cite{Poskanzer:1966}, it is fair to say that the spectrum of ${}^{15}$B has not been thoroughly understood yet. Experimentally, ${}^{15}$B has been studied using the in-beam $\gamma$-spectroscopy technique \cite{Stanoiu:2004} and multi-nucleon transfer reaction \cite{Kalpakchieva:2000}. Two bound excited states at $E_x=1.327$, $2.734$ MeV are observed in the $\gamma$-ray spectrum, and five excited states at $E_x=3.48(6)$, $4.90(6)$, $5.95(8)$, $7.63(8)$, $10.25(8)$ MeV in the ${}^{13}\text{C}({}^{14}\text{C},{}^{12}\text{N}){}^{15}\text{B}$ reaction, but spins/parities of these states have not been determined experimentally yet. { Recently, there are also many effects made to measure reaction cross sections for $^{13}$B, $^{14}$B, and $^{15}$B on different targets \cite{Estrade:2014aba,Tanaka:2017dow}.} On the theoretical side, at least four influential predictions could be found in literature, including three shell-model calculations \cite{Poppelier:1985,Warburton:1992rh,Stanoiu:2004} and one antisymmetrized-molecular-dynamics (AMD) calculation \cite{KanadaEnyo:1995ir}. In these theoretical studies, properties of the excited state at $E_x=3.48(6)$ MeV are less clear, and different models predict different energies and spin/parity assignments.

%\begin{figure}[tb]
%\centering
%\begin{center}
%\includegraphics[scale=0.22]{EnergyLevel.pdf}
%\figcaption{\label{EnergyLevel} Energy Levels of ${}^{15}$B obtained in various experimental and theoretical studies, with all energies in the unit of MeV. Here we compare theoretical predictions based on the PWG shell model, WBT shell model, $\text{WBT}^*$ shell model, and AMD model.}
%\end{center}
%%\end{figure}

In this letter, we would like to help clarify the properties of the excited state at $3.48(6)$ MeV, which is just beneath the two-neutron disintegration threshold and is probably a three-body bound state. Thanks to the large hierarchy between the neutron separation energies of ${}^{13}$B, ${}^{14}$B, and ${}^{15}$B, it is plausible that ${}^{15}$B could also be studied using a three-body model made of an inert ${}^{13}$B core and two valence neutrons. These neutron separation energies could be found in Table \ref{BIsotrope}, and indeed, we have
\begin{align}
S_n({}^{15}\text{B}),S_n({}^{14}\text{B})\ll S_n({}^{13}\text{B}).
\end{align}
Similar hierarchical structures of nucleon separation energies could also be found in other three-body nuclei, such as two-neutron nuclei ${}^{6}$He, ${}^{11}$Li \cite{Zhukov:1993aw,Brida:2006ar}, ${}^{12}$Be \cite{Nunes:1996cyo,RomeroRedondo:2008ev}, ${}^{14}$Be \cite{Tarutina:2004cpy}, ${}^{17}$B \cite{Ren:1990xza}, ${}^{22}$C \cite{Horiuchi:2006ds}, ${}^{23}$N \cite{Zhang:2014lzv}, and two-proton nuclei ${}^{17}$Ne \cite{Zhukov:1995zz,Garrido:2003jh}, ${}^{18}$Ne, ${}^{28}$S \cite{Chu:2008}, etc, and are viewed as clues of internal three-body structures. For other interesting discussions on nuclear three-body systems, see, for example, Ref.~\cite{Federov:1994cf,Ren:1994zz,Ren:1996,Lin:2012}.

%\begin{table}[t]
\begin{center}
\tabcaption{\label{BIsotrope} Physical properties of ${}^{13}$B, ${}^{14}$B and ${}^{15}$B. All energy scales are in the unit of MeV, and all lengths are in the unit of fm. The experimental data are taken from Ref.~\cite{NNDC} unless otherwise noted. Experimental errors are shown in brackets. Data in square brackets are based on theoretical expectations rather than experimental measurements.} 
%\centering
\footnotesize
\begin{tabular*}{80mm}{c@{\extracolsep{\fill}} cccc}
\\[-1.5ex]
\toprule
 ${}^{\text{A}}$X & $J^\pi$ & $T$ & $S_n$ & $S_{2n}$ \\[0.5ex] \hline\\[-2.5ex]
 ${}^{13}$B & $3/2^-$ & [3/2] & 4.879(17) & 8.248(10)  \\[0.5ex]
 ${}^{14}$B & $2^-$ & [2] & 0.970(21) & 5.848(21)  \\[0.5ex]
 ${}^{15}$B & $3/2^-$ \cite{Sauvan:2000hq} & [5/2] & 2.780(3) & 3.747(21)  \\
 \bottomrule
%\end{longtable}
\end{tabular*}
%\end{table}
\end{center}

The rest part of this note is organized as follows: In Section 2, we review briefly the Faddeev formalism that is used to solve the three-body Schr\"odinger equation, and introduce the corresponding interaction models. In Section 3, we present the numerical results concerning the two-body and three-body calculations. We end this note with conclusions in Section 4.

\section{Faddeev Formalism and Interaction Models}

{
In the present work, we solve the three-body wave functions using the Faddeev formalism proposed in Ref.~\cite{Thompson:2004dc},
\begin{align}
&\Psi^{JM}=\Psi^{JM}_1(\mathbf{x}_1,\mathbf{y}_1)+\Psi^{JM}_2(\mathbf{x}_2,\mathbf{y}_2)+\Psi^{JM}_3(\mathbf{x}_3,\mathbf{y}_3),\nonumber\\
%\end{align}
%\begin{align}
&(T_1+V_{3b}-E)\Psi^{JM}_1=-V_{23}(\Psi^{JM}_1+\Psi^{JM}_2+\Psi^{JM}_3),\nonumber\\
%\end{align}
%\begin{align}
&(T_2+V_{3b}-E)\Psi^{JM}_2=-V_{13}(\Psi^{JM}_1+\Psi^{JM}_2+\Psi^{JM}_3),\nonumber\\
%\end{align}
%\begin{align}
&(T_3+V_{3b}-E)\Psi^{JM}_3=-V_{12}(\Psi^{JM}_1+\Psi^{JM}_2+\Psi^{JM}_3),\label{FE3}
\end{align}
with $(\mathbf{x}_i,\mathbf{y}_i)$ one of the three Jacobi coordinate systems, 
\begin{align}
&\mathbf{x}_i=\sqrt{A_{jk}}\mathbf{r}_{jk},\quad \mathbf{y}_i=\sqrt{A_{i,jk}}\mathbf{r}_{i,jk},\nonumber\\
&\mathbf{r}_{jk}=\mathbf{r}_j\!-\!\mathbf{r}_k,\quad \mathbf{r}_{i,jk}=\mathbf{r}_i\!-\!(A_j\mathbf{r}_j\!+\!A_k\mathbf{r}_k)/(A_j\!+\!A_k),\nonumber\\
&A_{jk}=A_jA_k/(A_j+A_k),\nonumber\\
&A_{i,jk}=(A_j+A_k)A_i/(A_i+A_j+A_k).\nonumber
\end{align}
$T_i=T_{xi}+T_{yi}$ is the corresponding relative kinetic energy, and $V_{jk}$ is the two-body interaction between clusters $j$ and $k$, and $V_{3b}$ is the three-body force introduced to take into account all those effects that go beyond the two-body interactions.} The indices $(i,j,k)$ pick their values in $(1,2,3)$ cyclically. The Faddeev equations Eq.~\eqref{FE3} are solved by first transforming from the Jacobi coordinate system to the hyperspherical coordinate system
\begin{align}
&(\mathbf{x}_i,\mathbf{y}_i,s_{j},s_{k},I)\equiv(x_i,y_i,\Omega_{x_i},\Omega_{y_i},s_{j},s_{k},s_{i})\nonumber\\
\Longrightarrow & (\rho,\theta_i,\Omega_{x_i},\Omega_{y_i},s_j,s_k,s_i)\equiv (\rho,\Omega_{5i},s_j,s_k,s_i),\nonumber
\end{align} 
with $\rho^2=x^2_i+y^2_i$ the hyperradius, and $\theta_i=\arctan(x_i/y_i)$ the hyperangle. $(s_j,s_k,s_i)$ are the spins. We then introduce the hyperharmonic functions of $(\Omega_{5i},s_j,s_k,s_i)$, excluding only the dependence on $\rho$,
\begin{align}
&\mathcal{Y}^{L_iS_iJ_is_i,JM}_{K_il_{x_i}l_{y_i}}(\Omega_{5i},s_j,s_k,s_i)=\varphi^{l_{x_i}l_{y_i}}_{K_i}(\theta_i)\Big\{\big([Y_{l_{x_i}}\!(\Omega_{x_i})\nonumber\\
&\otimes Y_{l_{y_i}}\!(\Omega_{y_i})]_{L_i}\!\!\otimes\left[X_{s_j}\otimes X_{s_k}\right]_{S_i}\!\big)_{J_i}\otimes X_{s_i}\Big\}_{JM},\nonumber
\end{align}
with
\begin{align}
\varphi^{l_{x_i}l_{y_i}}_{K_i}(\theta_i)&=\mathcal{N}^{l_{x_i}l_{y_i}}_{K_i}(\sin\theta_i)^{l_{x_i}}(\cos\theta_i)^{l_{y_i}}\nonumber\\
&\times P^{l_{x_i}+1/2,l_{y_i}+1/2}_{n_i}(\cos2\theta_i).\nonumber
\end{align}
Here, $(l_{x_i},l_{y_i})$ are the orbital angular momenta. $Y_{l_{x_i}m_{x_i}}(\Omega_{x_i})$ and $X_{s_i}$ are the spherical and spin harmonics, respectively, with square brackets being the standard Clebsch-Gordan combination of two angular momenta. $P^{l_{x_i}+1/2,l_{y_i}+1/2}_{n_i}(\cos2\theta_i)$ are the Jacobi polynomials. $K_i=l_{x_i}+l_{y_i}+2n_i\ (n_i=0,1,2,\cdots)$ is the hyper-angular-momentum. The normalization constant $\mathcal{N}^{l_{x_i}l_{y_i}}_{K_i}$ is given by
\begin{align}
\mathcal{N}^{l_{x_i}l_{y_i}}_{K_i}=\left[\frac{2(K_i+2)\Gamma(K_i+2-n_i)\Gamma(n_i+1)}{\Gamma(n_i+l_{x_i}+3/2)\Gamma(n_i+l_{y_i}+3/2)}\right]^{1/2}.\nonumber
\end{align}
With the help of hyperharmonic functions, the Faddeev components $\Psi^{JM}_i$ could be rewritten as
\begin{align}
\Psi^{JM}_i(\mathbf{x}_i,\mathbf{y}_i)&=\rho^{-5/2}\!\!\!\!\!\!\!\!\!\!\!\!\!\!\sum_{l_{x_i}l_{y_i}L_iS_iJ_is_i,K_i}\!\!\!\!\!\!\!\!\!\!\!\!\!\!\mathcal{X}^{L_iS_iJ_is_iJ}_{i,K_il_{x_i}l_{y_i}}(\rho)\nonumber\\
&\times\mathcal{Y}^{L_iS_iJ_is_i,JM}_{K_il_{x_i}l_{y_i}}(\Omega_{5i},s_j,s_k,s_i).
\label{HHE}
\end{align} 
After inserting Eq.~\eqref{HHE} into the Faddeev equations, one obtains a set of coupled ordinary differential equations for $\mathcal{X}^{L_iS_iJ_is_iJ}_{i,K_il_{x_i}l_{y_i}}(\rho)$, which could be solved using the modern numerical algorithms for differential equations. We recommend Ref.~\cite{Thompson:2004dc} for a detailed discussion on the implementation of the Faddeev formalism including the three-body forces. 

To predict physical properties of ${}^{15}$B quantitatively, we need to determine first the neutron-neutron and neutron-core interaction models by fitting experimental data. Here, we assume that the ${}^{13}$B ground state has the neutron configuration $(0s_{1/2})^2(0p_{3/2})^4(0p_{1/2})^2$. The ground state, first and second excited states of ${}^{14}$B are assumed to have the neutron configurations $(0s_{1/2})^2$$(0p_{3/2})^4$$(0p_{1/2})^2$$(1s_{1/2})^1$, $(0s_{1/2})^2$$ (0p_{3/2})^4(0p_{1/2})^2(1s_{1/2})^1$, and $(0s_{1/2})^2$ $(0p_{3/2})^4$$(0p_{1/2})^2$ $(0d_{5/2})^1$, respectively, which are consistent with the Nordheim weak rule for the odd-odd nuclei \cite{Nordheim:1950zz}, as well as the explicit shell-model calculations in Ref.~\cite{Bedoor:2016vzb}. Noticeably, the ground state and the first excited state of ${}^{14}$B are assumed to have the same neutron configuration, and the splitting in their energies corresponds to the hyperfine structure resulted from the spin-spin interaction between the ${}^{13}$B core and the valence neutron.

 For the neutron-neutron interaction, we adopt the Gogny-Pires-Tourreil (GPT) potential \cite{Gogny:1970jbc}. For the neutron-core interaction, we adopt a Gaussian form with the spin-dependent interaction given by the Garrido-Fedorov-Jensen (GFJ) ansatz \cite{Garrido:2003ma,Garrido:2003jh},
\begin{align}
V_\text{n-core}(r)&=\exp(-r^2/b^2)(V_\text{C}+V_\text{SO}\,\mathbf{L}\cdot\mathbf{s}_n\nonumber\\
&+V_\text{SS}\,\mathbf{J}_n\!\cdot\mathbf{s}_c),
\label{NeutronCoreInteraction}
\end{align}
with $\mathbf{L}$ the orbital angular momentum, $\mathbf{s}_n$ the spin of the valence neutron, $\mathbf{s}_c$ the spin of the ${}^{13}$B core, and $\mathbf{J}_n=\mathbf{s}_n+\mathbf{L}$. Naively, there could be other choices for spin-dependent interactions, like
\begin{align}
 &V_\text{SS}\,\mathbf{s}_n\cdot\mathbf{s}_c+V_{\text{SO}_n}\,\mathbf{L}\cdot\mathbf{s}_n+V_{\text{SO}_c}\,\mathbf{L}\cdot\mathbf{s}_c,\nonumber\\
 &V_\text{SS}\,\mathbf{s}_n\cdot\mathbf{s}_c+V_\text{SO}\,\mathbf{L}\cdot(\mathbf{s}_n+\mathbf{s}_c),\nonumber  \\
 &V_\text{SS}\,\mathbf{s}_n\cdot\mathbf{s}_c+V_\text{SO}\,\mathbf{L}\cdot\mathbf{s}_n,\ \text{etc}.\nonumber
\end{align}
However, Ref.~\cite{Garrido:2003ma} shows that choices different from Eq.~\eqref{NeutronCoreInteraction} would probably lead to wrong predictions on the excited states, even if the ground-state properties are forced to be reproduced exactly. 

Free parameters $b$, $V_\text{C}$, $V_\text{SO}$, and $V_\text{SS}$ in Eq.~\eqref{NeutronCoreInteraction} are determined by reproducing the following conditions: the root mean square (RMS) radius of ${}^{13}$B $r_0=1.23A^{1/3}=2.89$ fm; the ground state $2^-$ of ${}^{14}$B with the valence neutron in the $1s_{1/2}$ orbit has the energy $\epsilon[1s_{1/2}(2^-)]=-0.970$ MeV beneath the ${}^{13}\text{B}+n$ threshold \cite{NNDC}; the first excited state $1^-$ of ${}^{14}$B with the valence neutron in the $1s_{1/2}$ orbit has the energy of $\epsilon[1s_{1/2}(1^-)]=-0.316$ MeV beneath the ${}^{13}\text{B}+n$ threshold \cite{Kanungo:2005rxc}; the last neutron in the ${}^{13}$B ground state $3/2^-$ occupies the $0p_{1/2}$ orbit, and its energy could be estimated by the neutron separation energy of ${}^{13}$B, i.e., $\epsilon[0p_{1/2}(1^-)]=-4.879$ MeV \cite{NNDC}; the excited state $3^-$ of ${}^{14}$B with the valence neutron in the $0d_{5/2}$ orbit has the resonance energy $\epsilon[0d_{5/2}(3^-)]=0.41$ MeV above the ${}^{13}\text{B}+n$ threshold \cite{NNDC}. $b$, $V_\text{C}$, $V_\text{SO}$, and $V_\text{SS}$ determined thereby are summarized in Table \ref{VcnFreeParameters}. We also need to introduce a three-body interaction of Gaussian type to account for deviations between computed results with only bare two-body interactions and experimental data, as well as to simulate effects of core deformations and/or core excitations \cite{RomeroRedondo:2008ev},
\begin{align}
V_{3b}(\rho)=V_\text{3B}\exp\!\!\left[-\!\!\left(\frac{\rho}{r_\text{3B}}\right)^2\right].
\end{align}
Explicitly, three three-body interaction parameter sets are considered in this work (see Table \ref{ThreeBodyModel}).  

\vspace{0.5mm}

%\begin{table}[t]
\begin{center}
\tabcaption{\label{VcnFreeParameters} Interaction parameters for the neutron-core interaction used in this work. All lengths are in the unit of fm, while all energies are in the unit of MeV.} 
%\centering
\begin{tabular*}{80mm}{c@{\extracolsep{\fill}}ccccc}
%{p{1.4cm}p{0.5cm}p{0.8cm}p{1.2cm}p{1.2cm}p{0.5cm}}
\\[-1.5ex]
%\hline\hline\\[-2.5ex]
\toprule
Parameter & $b$ & $V_\text{C}$ & $V_\text{SO}^{(l\neq2)}$ & $V_\text{SO}^{(l=2)}$ &  $V_\text{SS}$ \\[0.5ex] \hline\\[-2.5ex]
Value & $2.28$ & $-87.4$ & $-13.65$ & $-27.165$ & $-3$  \\ 
%\hline\hline
\bottomrule
\end{tabular*}
\end{center}
%\end{table}

%\begin{table}[t]
\begin{center}
\tabcaption{\label{ThreeBodyModel} Three-body interaction parameter sets used in this work, with all lengths in the unit of fm, and all energies in the unit of MeV.} 
%\centering
\begin{tabular*}{80mm}{c@{\extracolsep{\fill}}ccc}
\\[-1.5ex]
%\hline\hline\\[-2.5ex]
\toprule
Parameter & V3BA & V3BB & V3BC \\[0.5ex] \hline\\[-2.5ex]
 $r_\text{3B}$ & 4.0 & 5.0 & 6.0  \\[0.5ex]
 $V_\text{3B}$ & -5.69 & -3.44 & -2.54 \\ 
 %\hline\hline
 \bottomrule
\end{tabular*}
\end{center}
%\end{table}

%\vspace{0.5mm}

%\begin{table}[t]
\begin{center}
\tabcaption{{ \label{SingleParticleStateB14} RMS matter radii and energies of various states in ${}^{13}$B and ${}^{14}$B obtained by parameters in Table \ref{VcnFreeParameters}, with all lengths in the unit of fm, and all energies in the unit of MeV. $r_m(^{14}\text{B})\equiv r_\text{m}[1s_{1/2}(2^-)]$ is the RMS matter radii for the $^{13}$B ground state, while $r_m(^{13}\text{B})\equiv r_\text{m}[0p_{1/2}(1^-)]$ is the RMS matter radii for the $^{14}$B ground state.}} 
%\centering
\footnotesize
\begin{tabular*}{80mm}{c@{\extracolsep{\fill}}cc}
%\hline\hline\\[-2.5ex]
\toprule
$\!\!r_m(^{13}\text{B})$ & $\!\!r_m(^{14}\text{B})$ & $\!\!r_\text{m}[1s_{1/2}(1^-)]$ \\[0.5ex] \hline\\[-2.5ex] 
\\[-2.5ex]
$2.88$ & $3.14$ & $3.51$\\
% \hline\hline
\bottomrule
\end{tabular*}
\begin{tabular*}{80mm}{c@{\extracolsep{\fill}}ccc}
%{p{1.48cm}p{1.48cm}p{1.48cm}p{1.48cm}}
\\
%\hline\hline\\[-2.5ex]
\toprule
$\!\!\epsilon[0p_{1/2}(1^-)]$ & $\ \ \ \epsilon[1s_{1/2}(2^-)]$ & $\!\!\epsilon[1s_{1/2}(1^-)]$ & $\!\!\epsilon[0d_{5/2}(3^-)]$ \\[0.5ex] \hline\\[-2.5ex]
$-4.919$ & $-0.917$ & $-0.316$ & $0.410$ \\ 
%\hline\hline\\
\bottomrule
\end{tabular*}
\end{center}
%\end{table}

\section{Numerical Results}

Numerical results for the two-body and three-body calculations are discussed as follows. With interaction parameters in Table \ref{ThreeBodyModel}, { RMS matter radii and single-particle energies of various ${}^{13}$B and ${}^{14}$B states could be found in Table \ref{SingleParticleStateB14}. The valence-neutron radius and the total matter radius of the ${}^{14}$B ground state are given by $r_\text{n}=5.51$ fm and $r_\text{m}[1s_{1/2}(2^-)]=3.14$ fm, respectively.} The matter radius turns out to be a bit larger than the naive estimation $r_0=1.23A^{1/3}=2.94$ fm, revealing the existence of neutron halo in the ${}^{14}$B ground state. Similar arguments could also be applied to the first excited state $2^-$. The density distributions of the valence neutron in various ${}^{13}$B and ${}^{14}$B states are shown in Fig.~\ref{SingleNeutronEnergy}. The resonance state $3^-$ of ${}^{14}$B is located by calculating the two-body scattering process of the valence neutron and ${}^{13}$B and determining its resonant peak in the energy curve (see Fig.~\ref{SingleNeutronResonance}).

%\begin{figure}[tb]
%\centering
\begin{center}
\includegraphics[width=0.45\textwidth]{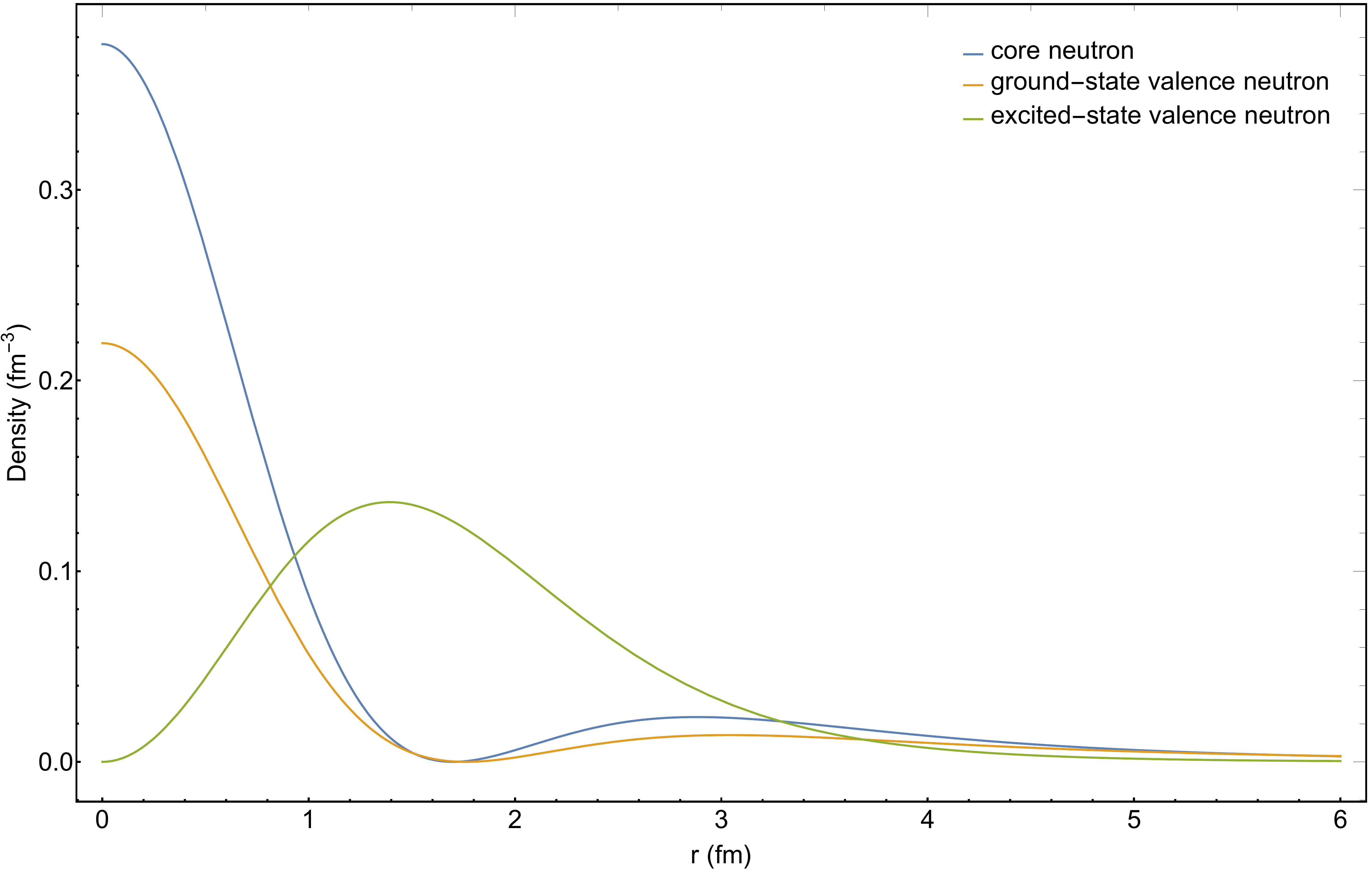}
\figcaption{{The density distribution of neutrons in the subsystem $^{14}$B. The blue curve is the density distribution of the core neutron in ${}^{13}$B and the orange and green curves are the neutron density distributions of the valence neutron in the ground state and first excited state of ${}^{14}$B, respectively.}}
\label{SingleNeutronEnergy} 
\end{center}
%\end{figure}

%\begin{figure}[tb]
\begin{center}
%\centering
\includegraphics[width=0.45\textwidth]{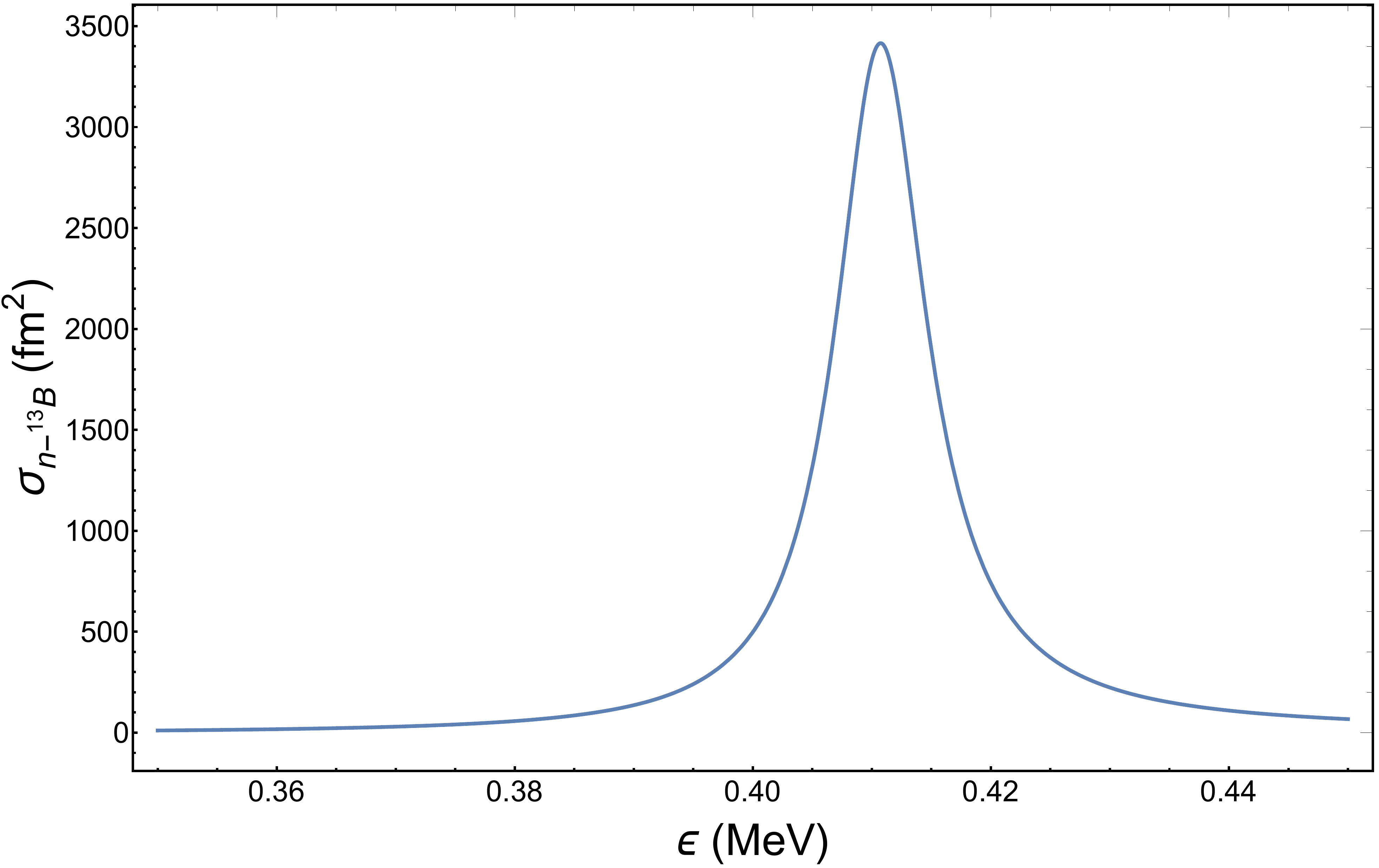}
\figcaption{Two-body cross section of the valence neutron and ${}^{13}$B as a function of the two-body total energy $\epsilon$. The resonance state $3^-$ corresponds to the peak in the energy curve.}
\label{SingleNeutronResonance} 
%\end{figure}
\end{center}

In the three-body calculation of ${}^{15}$B, one has to first carry out the so-called Pauli blocking procedure to remove the deeply bound states of the neutron-core two-body interaction, which are assumed to be occupied by core neutrons in ${}^{13}$B already. This is done through a supersymmetry transformation \cite{Baye:1987,Baye:1994,Sparenberg:1997}. In our calculation, we use heavily the hyperspherical harmonics expansion, and it is important to check out its convergence property. The truncation of the hyperspherical harmonics expansion is controlled by the hypermomentum $K$. {In Fig.~\ref{Convergence}, we show the convergence of the ${}^{15}$B ground state energy $E_\text{gs}$ in the three-body interaction parameter set V3BA as $K_\text{max}$ increases. The V3BB and V3BC parameter sets show similar convergence behavior and will not be discussed explicitly here. One could see intuitively that for $K_\text{max}\sim20$, hyperspherical harmonics expansion has already show good convergence behavior for the ground-state energy. A similar check has also been carried out for the ${3/2}_2^-$ state. We find that the energy of the $3/2^-_2$ state decreases gradually as $K_\text{max}$ increases, affected a bit more significantly by the $K$ truncation compared with the ground-state energy. Therefore, in our calculation, we take $K_\text{max}=22$ to do our best to relieve the impacts of the $K$ truncation, which is also consistent with other calculations on $\text{core}+n+n$ systems (see, e.g., \cite{Nunes:1996cyo,Zhang:2014lzv}).} 

We calculate energies of the ground state ${3/2}_1^-$ and the excited state $3/2_2^-$ (with respect to $n+{}^{13}$B threshold), along with their RMS matter radii, using the two-body and three-body interaction models in Table \ref{VcnFreeParameters} and \ref{ThreeBodyModel}. The results are listed in Table \ref{ThreeBodyResults} with all energies given with respect to the ${}^{13}\text{B}+n+n$ threshold, and are numerically close to each other, revealing encouraging robustness of our predictions. The three-body interaction parameter sets in Table \ref{ThreeBodyModel} are chosen to reproduce the ground state energy of ${}^{15}$B $E_\text{gs}=-3.747$ MeV exactly. The RMS radius of the ground state $3/2_1^-$ is found to be about $3.085\sim3.104$ fm, corresponding to the size of a stable nucleus with $A\approx16$. In other words, the ground state $3/2_1^-$ is not a halo nucleus, which is consistent with the large two-neutron separation energy. { To see explicitly the effects of the three-body interaction, we also do the calculation by switching off the three-body interaction for comparison, and find that, in this case, the ground-state energy turns out to be $-2.696$ MeV. This shows that the three-body interaction indeed plays an important role in our calculations.}

%\begin{figure}[tb]
%\centering
\begin{center}
\includegraphics[width=0.45\textwidth]{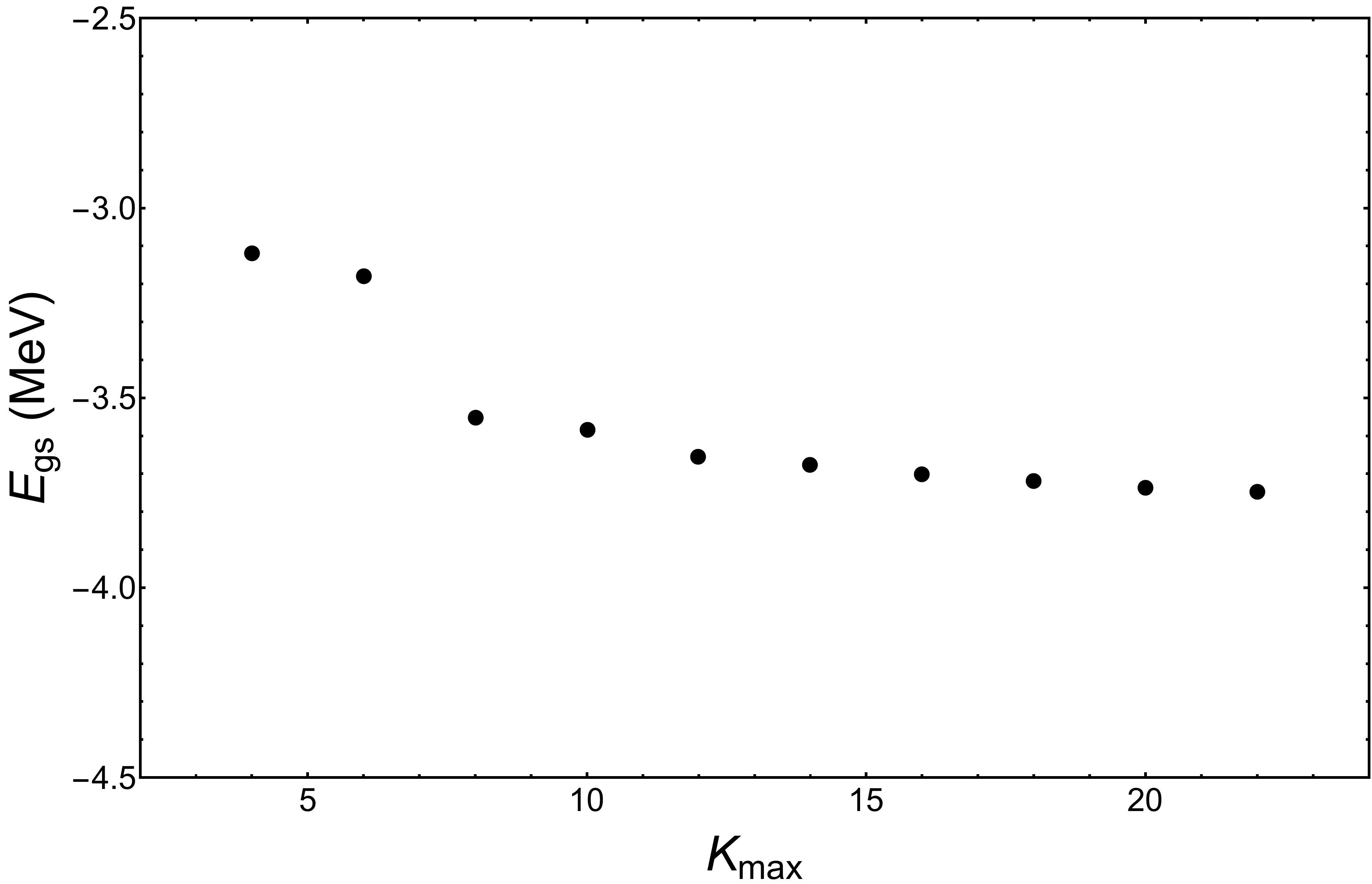}
\figcaption{Convergence of the ${}^{15}$B ground state energy as a function of the maximum $K$ value in the hyperspherical harmonics expansion.}
\label{Convergence} 
%\end{figure}
\end{center}

%\begin{figure}[tb]
%\centering
%\begin{center}
%\includegraphics[width=0.45\textwidth]{Convergence2.pdf}
%\figcaption{Convergence of the energy of the excited state ${3/2}^-_2$ as a function of the maximum $K$ value in the hyperspherical harmonics expansion.}
%\label{Convergence2} 
%%\end{figure}
%\end{center}

%\begin{table}[t]
\begin{center}
\tabcaption{Energies and RMS radii of the ground state $3/2_1^-$ and the excited state $3/2^-_2$, calculated with three different three-body interaction parameter sets. Quantities with $*$ correspond to the excited state. All energies are in the unit of MeV, while all lengths in the unit of fm.} 
\label{ThreeBodyResults}
%\centering
\begin{tabular*}{80mm}{c@{\extracolsep{\fill}}cccccc}
\\[-1.5ex]
%\hline\hline\\[-2.5ex]
\toprule
 & $E_\text{gs}$ & $r_\text{m}$ & $E^*$ & $r_\text{m}^*$  \\[0.5ex] \hline\\[-2.5ex]
 V3BA & -3.7475 & 3.085 & -0.1293 & 4.770 \\[0.5ex]
 V3BB & -3.7470 & 3.095 & -0.1320 & 4.763 \\[0.5ex]
 V3BC & -3.7465 & 3.104 & -0.1481 & 4.739 \\[0.5ex]
 Exp. & $-3.747(21)$ & $-$ & -0.267(64) & $-$ \\ 
 %\hline\hline
 \bottomrule
\end{tabular*}
%\end{table}
\end{center}

The excited state $3/2_2^-$, on the other hand, is quite interesting. First, its excitation energy is about $E^*=3.598\sim3.615$ MeV, which is energetically close to the third excited state observed experimentally at $E_x=3.48(6)$ MeV. It is thus plausible to identify these two states. This identification is consistent with the WBT and $\text{WBT}^*$ shell model calculations, which predict that the third excited state has spin/parity $3/2^-$ as well. The tiny difference between our three-body calculation and the experimental measurement may be a result of unresolved experimental errors or theoretical defects in model building. In the following discussions we shall ignore the difference between theoretical predictions and experimental measurements for simplicity.

Second, the tiny energy and the large RMS matter radius of the ${3/2}^{-}_2$ state indicate that it has a two-neutron halo. Indeed, the matter radius $r_m^*\approx4.763$ fm corresponds to the size of a stable nucleus with $A\approx58$.

To illustrate the inner structure of the $3/2^-_2$ state further, we calculate the RMS distance between two valence neutrons and that between the ${}^{13}$B core and the center of mass of the valence neutron pair, denoted by $r_\text{nn}$ and $r_\text{c,nn}$, respectively. The numerical results could be found in Table \ref{InnerStructureB15}. It is interesting to note that
%\begin{align}
$\frac{r_\text{nn}}{r_\text{c,nn}}\approx\frac{r_\text{nn}^*}{r_\text{c,nn}^*}\approx1.86$.
%\label{Similarity}
%\end{align}

%\begin{table}[t]
\begin{center}
\tabcaption{RMS distances between two valence neutrons $r_\text{nn}$ and from the ${}^{13}$B core to the valence neutron pair $r_\text{c,nn}$ in the ${}^{15}$B ground state ${3/2}^-_1$ and the excited state ${3/2}^-_2$. Quantities for the excited state are superscribed by an extra $*$. All lengths are in the unit of fm.} 
\label{InnerStructureB15}
%\centering
\begin{tabular*}{80mm}{c@{\extracolsep{\fill}}cccccc}
\\[-1.5ex]
%\hline\hline\\[-2.5ex]
\toprule
 & $r_\text{nn}$ & $r_\text{c,nn}$ & $r_\text{nn}^*$ & $r_\text{c,nn}^*$  \\[0.5ex] \hline\\[-2.5ex]
 V3BA & 5.86 & 3.15 & 15.26 & 8.20 \\[0.5ex]
 V3BB & 5.94 & 3.19 & 15.23 & 8.18 \\[0.5ex]
 V3BC & 6.00 & 3.23 & 15.12 & 8.12 \\ 
 %\hline\hline
 \bottomrule
\end{tabular*}
\end{center}
%\end{table}

The Spatial distributions of the valence neutrons could be calculated by
\begin{align}
P(r_\text{nn},r_\text{c,nn})&\equiv r_\text{nn}^2r_\text{c,nn}^2\nonumber\\
&\times\int\left|\Psi^{JM}(\mathbf{r}_\text{nn},\mathbf{r}_\text{c,nn})\right|^2\mathrm{d}\Omega_{\mathbf{r}_\text{nn}}\mathrm{d}\Omega_{\mathbf{r}_\text{c,nn}},\nonumber
\end{align}
which are displayed pictorially in Fig.~\ref{SpatialDistribution4GS} and \ref{SpatialDistribution4EX} with the three-body interaction parameter set V3BA. For the ground state ${3/2}_1^-$, the spatial distribution function $P(r_\text{nn},r_\text{c,nn})$ is peaked at around $(r_\text{nn},r_\text{c,nn})=(4.4,2.4)$ fm, while for the excited state ${3/2}_2^-$, $P(r^*_\text{nn},r^*_\text{c,nn})$ has a global maximum at around $(12.5,6.7)$ fm, as well as a secondary maximum at around $(4.1,2.2)$ fm. The dominant occupation probabilities of valence neutrons for the ground state and $3/2_2^-$ excited state are given in Table \ref{Occupation}, with three different three-body interaction parameter sets. Once again, one observes excellent convergence of numerical results among different interaction parameter sets.

%\begin{figure}[tb]
\begin{center}
%\centering
\includegraphics[width=0.45\textwidth]{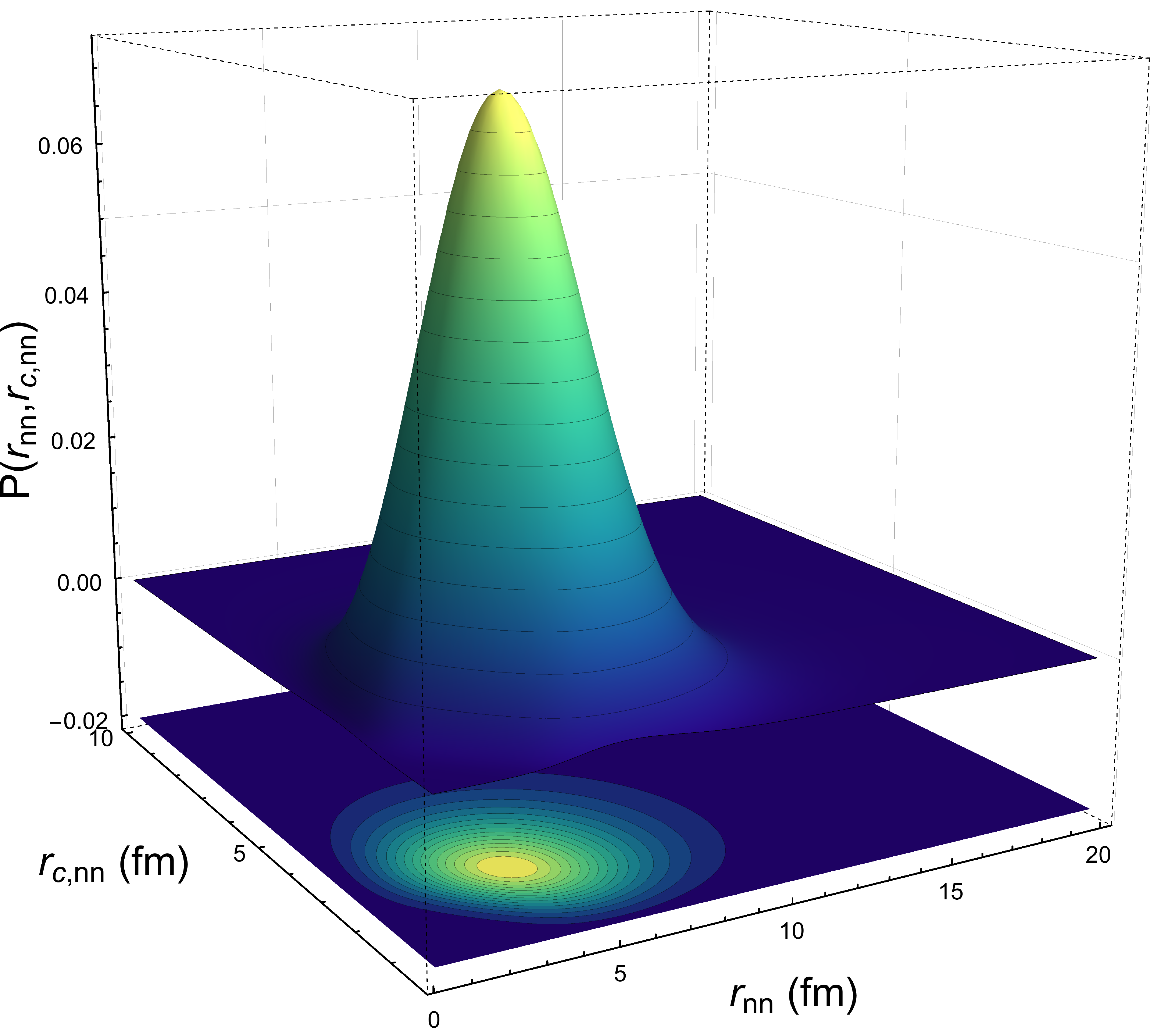}
\figcaption{Spatial distribution of two valence neutrons in the ground state ${3/2}^-_1$ of ${}^{15}$B with the three-body interaction parameter set V3BA.}
\label{SpatialDistribution4GS} 
%\end{figure}
\end{center}

%\begin{figure}[tb]
\begin{center}
%\centering
\includegraphics[width=0.45\textwidth]{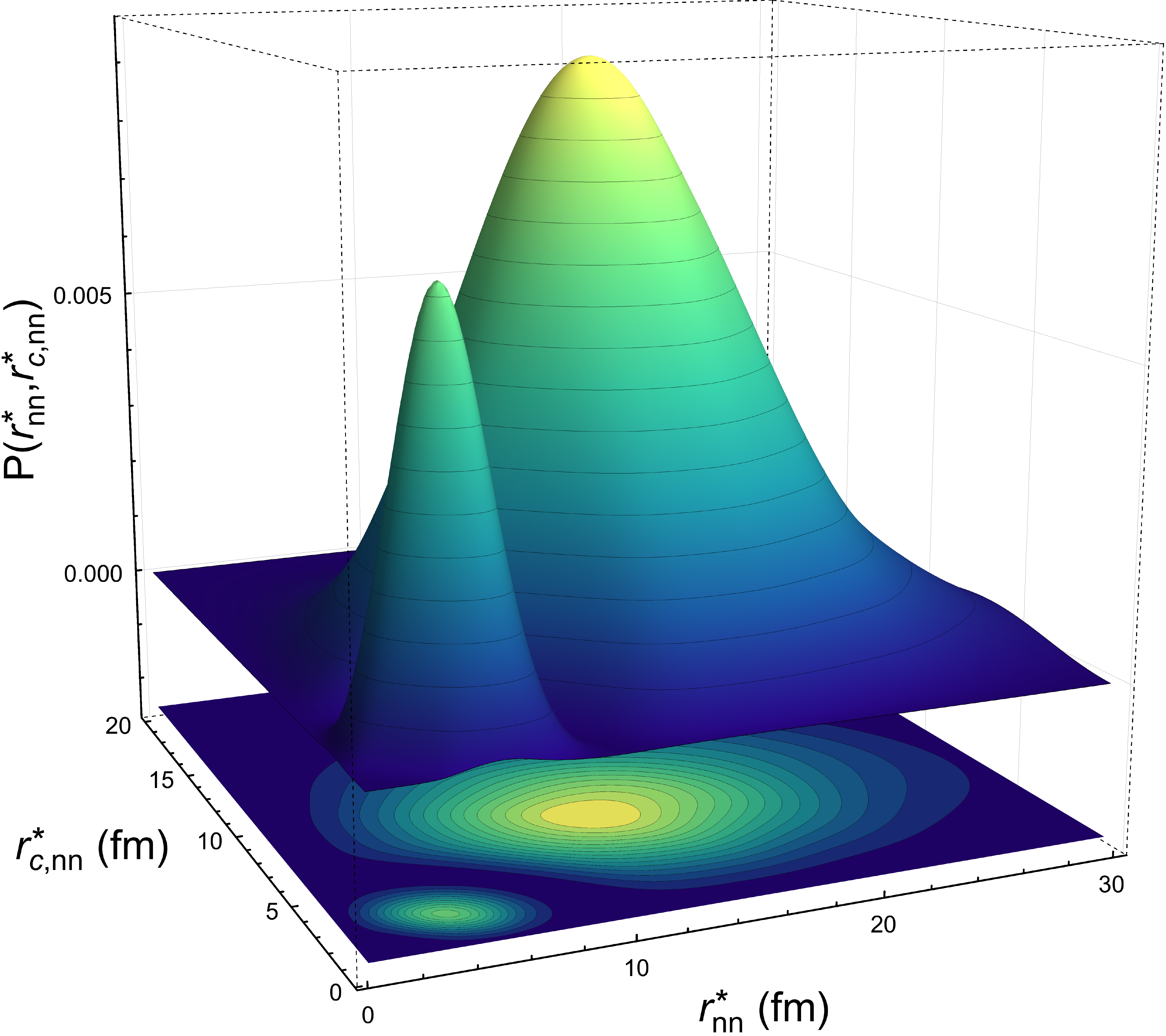}
\figcaption{Spatial distribution of two valence neutrons in the excited state ${3/2}^-_2$ of ${}^{15}$B with the three-body interaction parameter set V3BA.}
\label{SpatialDistribution4EX} 
%\end{figure}
\end{center}

%\begin{table}[t]
\begin{center}
\tabcaption{Dominant occupation probabilities of valence neutrons for the ground state and ${3/2}^-_2$ excited state of ${}^{15}$B, calculated using three different three-body interaction parameter sets. } 
\label{Occupation}
%\centering

\begin{tabular*}{80mm}{c@{\extracolsep{\fill}}cccccc}
\\[-1.5ex]
%\hline\hline\\[-2.5ex]
\toprule
$3/2^-_1$ & V3BA &  V3BB & V3BC  \\[0.5ex] \hline\\[-2.5ex]
$(1s_{1/2})^2$ & 96.10\% & 96.25\% & 96.28\%   \\[0.5ex]
$(0d_{5/2})^2$ & 2.10\% & 2.01\% &  1.98\% \\[0.5ex]
$(0d_{3/2})^2$ & 1.40\% & 1.33\% &  1.32\%  \\ 
%\hline\hline
\bottomrule
\end{tabular*}

\begin{tabular*}{80mm}{c@{\extracolsep{\fill}}cccccc}
\\[-1.5ex]
%\hline\hline\\[-2.5ex]
\toprule
$3/2^-_2$ & V3BA &  V3BB & V3BC  \\[0.5ex] \hline\\[-2.5ex]
$(1s_{1/2})^2$ & 86.32\% & 86.32\% &  86.36\%  \\[0.5ex]
$(0d_{5/2})^2$ & 7.23\% & 7.23\% &  7.21\%  \\[0.5ex]
$(0d_{3/2})^2$ & 4.82\% & 4.82\% &  4.81\%  \\ 
%\hline\hline
\bottomrule
\end{tabular*}

\end{center}
%\end{table}

\section{Conclusions}

In summary, in this letter we investigate the low-lying bound states of ${}^{15}$B by assuming it is a three-body system made of an inert ${}^{13}$B core and two valence neutrons. It is plausible to identify the excited state $3/2_2^-$ appearing in our model with the experimentally observed excited state at $E_x=3.48(6)$ MeV. Such an identification is also consistent with the WBT and $\text{WBT}^*$ shell model calculations. We then study in detail the properties of the ground state $3/2_1^-$ and the excited state $3/2_2^-$, calculating their energies with respect to the three-body disintegration threshold, RMS matter radii, wave functions, occupation probabilities, etc. Our calculations are carried out with three different three-body parameter sets, and numerical predictions show excellent convergence behavior. It is found that unlike the ground state $3/2^-_1$, the excited state $3/2_2^-$ has a giant two-neutron halo with the matter radius comparable to the size of a stable nucleus with $A\approx58$. The studies here might be helpful for deepening our understanding of dripline phenomenology.
\\

%\section*{Acknowledgements}

\acknowledgments{
%D.~B.~ would like to thank Prof.~I.J.~Thompson deeply for his helpful and patient communications. 
D.~B.~would like to thank Huabin Cai, Daming Deng, Bin Hong, Songju Lei, Hao Lu, Wan Niu, Xin-xing Shi and Xin Zhang for discussions. This work is supported by the National Natural Science Foundation of China (Grant No.~11535004, 11761161001, 11375086, 11120101005, 11175085 and 11235001), by the National Major State Basic Research and Development of China, Grant No.~2016YFE0129300, and by the Science and Technology Development Fund of Macau under Grant No.~068/2011/A.
}

\end{multicols}

\vspace{-1mm}
\centerline{\rule{80mm}{0.1pt}}
\vspace{2mm}

\begin{multicols}{2}

\end{multicols}

\clearpage

\end{CJK*}
\end{document}